# Sum Rate Capacity of MIMO HetNet Systems in the Presence of Channel Estimation Error

Esha Bangar
The University of Texas at Dallas
Richardson, TX 75093
Esha.bangar@utdallas.edu

Nima Taherkhani
The University of Texas at Dallas
Richardson, TX 75093
Nima.Taherkhani@utdallas.edu

Kamran Kiasaleh, *Senior Member*, *IEEE*
The University of Texas at Dallas
Richardson, TX 75093
kamran@utdallas.edu

*Abstract*— In this paper, the impact of channel estimation error (CEE) on the sum-rate capacity of multiple-input-multiple-output (MIMO) heterogeneous networks (HetNets) is investigated. It is assumed that the receiver is a linear minimum mean-square (LMMS) receiver. The architecture is based on the deployment of macro base stations with large antenna arrays and a secondary tier of small cell base stations having fewer number of antenna arrays. The key contribution of the paper is to highlight the noticeable impact of CEE on the sum-rate capacity of macro-cells. Furthermore, it is shown that CEE has only marginal impact on the sum-rate capacity of small cells. Simulations for the sum rate of macro users versus that of the small cell users for time-division duplex (TDD) are performed and the results are compared with the case when channel estimation error is present.

## I. INTRODUCTION

The immense increase in the demand for reliable, high-speed data services has led to three approaches to increase the capacity of wireless networks. One consideration assumes spatial diversity using a large number of antenna arrays, known as massive MIMO. The second approach considers the deployment of a combination of different tiers of cells, which is referred as heterogeneous networks, to enhance capacity. The third approach seeks to combine the previous two methods to achieve the highest level of flexibility and capacity in terms of bits/sec/Hz/Km$^2$ [1]-[8]. However, to take full advantage of the capacity of massive MIMO in a HetNet setting, accurate channel state information must be available at the transceiver.

In any practical system, perfect channel state information is not available at the receiver. In fact, very often, non-negligible CEE sets a lower bound on the system error rate (*i.e.,* error floors). Therefore, a key requirement of a successful deployment of HetNet systems with massive MIMO capability is perfect channel estimation. Perfect channel state information is usually assumed in the literature when analysing the performance of MIMO and HetNets using linear detectors. In [1], the authors have considered MIMO Hetnet architecture and compared the spectral efficiency of macro cell users and small cell users for uplink and downlink. In [2], the authors have proposed an inter-tier interference scheme for the MIMO Hetnets architecture but have made the impractical assumption of having perfect channel state information for the local channels at the base station (BS). CEE has been considered in [3], [4] and [5] for the case of orthogonal frequency-division multiplexed (OFDM) and MIMO architecture with MMSE receiver. However, these studies do not consider a HetNets setting. Therefore, the impact of CEE in the case of small-cells has not been investigated in the literature. In [6], authors have considered CEE for MIMO Hetnets architecture, but have made two major assumptions for the analysis. First, small-cell BSs are considered for backhaul only and all the users are served by the macro-cell BS. Second, the paper groups all the users together and studies the impact of CEE on their performances instead of making an attempt to measure the impact of CEE on users being served by different tiers. A similar argument is made in [7], where CEE is considered for MIMO Hetnets. This study also ignores the impact of CEE for the small-cells by stating that the small-cell channels can be perfectly estimated owing to the slow movement of the small-cell users. In this study, we demonstrate that, even though CEE has smaller impact on the performance of small-cell users as compared to that of the macro-cell users, the impact of CEE cannot be simply ignored for small-cell users.

In this paper, we present an architecture consisting of MIMO macro tier base stations overlaid with a secondary tier of small cells having fewer number of antennas. We analyse the performance of MMSE receivers in the presence of CEE for MIMO HetNet system and derive an expression for SINR which provides an estimate of the sum-rate performance of macro-cell users and small-cell users in the presence of CEE. The sum-rate performance is investigated for both cases when the channel is perfectly known at the transmitter and when imperfect channel state information is available for TDD duplexing scheme.

## II. SYSTEM MODEL

We are considering a two-tier network consisting of $B$ macro base stations (BS) wherein each cell is overlaid with $S$ small cells (SC). The BS's and SC's are equipped with $N$ and $F$ antennas, respectively. Each BS serves $K \leq N$ macro-cell users (MCU) and each SC serves a single small-cell user (SCU). The MCU's and SCU's are assumed to be single antenna users, as is done in practice.

The system architecture is described in Fig. 1. We consider that the two tiers operate in different frequency bands; however, their uplink and downlink streams are separated in time utilizing TDD scheme.

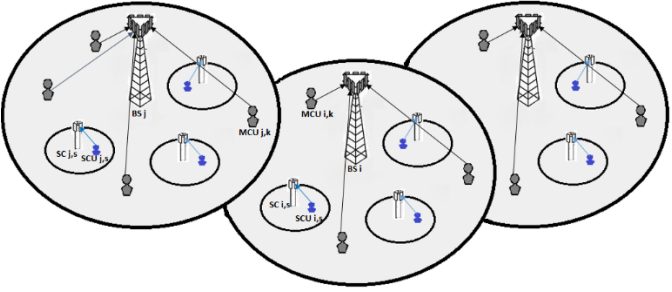

Fig. 1 – System Architecture with MIMO BSs and overlaid SCs

The received uplink $1 \times N$ signal vector at the $i^{th}$ BS at time $t$ is given by [1]:

$$\boldsymbol{y}_i^{BS}(t) = \sum_{b=1}^{B}\left(\sum_{k=1}^{K}\sqrt{P_{MCU}}\,\boldsymbol{h}_{ibk}^{BS-MCU}x_{bk}(t)\right) + \boldsymbol{n}_i(t) \quad (1)$$

where $\boldsymbol{h}_{ibk}^{BS-MCU}$ is the $1 \times N$ channel state vector between the $k^{th}$ MCU in cell $b$ and the BS $i$. $P_{MCU}$ and $P_{SCU}$ are the transmit powers of MCU's and SCU's, respectively. Furthermore, $x_{bk}$ is the transmit signal of MCU, while $\boldsymbol{n}_i(t)$ denotes the $1 \times N$ additive white Gaussian noise(AWGN) vector with zero mean and covariance matrix $N_0\,\boldsymbol{I}_N$ at BS $i$, where $\boldsymbol{I}_N$ is an $N \times N$ identity matrix.

The received uplink $1 \times F$ signal vector at the $j^{th}$ SC in the $i^{th}$ macro-cell at time $t$, denoted by $\boldsymbol{y}_{ij}^{SC}(t)$, can be expressed as [1]:

$$\boldsymbol{y}_{ij}^{SC}(t) = \sum_{b=1}^{B}\left(\sum_{s=1}^{S}\sqrt{P_{SCU}}\,\boldsymbol{h}_{ijbs}^{SC-SCU}x_{bs}(t)\right) + \boldsymbol{n}_{ij}(t) \quad (2)$$

where $\boldsymbol{h}_{ijbs}^{SC-SCU}$ denotes the $1 \times F$ channel vector between $s^{th}$ SCU in cell $b$ and the $j^{th}$ SC in the macro-cell $i$. $x_{bs}$ is the transmit signal of SCU. Furthermore, $\boldsymbol{n}_{ij}(t)$ denotes the $1 \times F$ AWGN vector for the reception of signals at the $j^{th}$ SC in the $i^{th}$ macro-cell.

In case of co-channel TDD, the received uplink signal vector at BS (i.e., (1)) will include the SCU signal terms as the SCUs will be operating in the same frequency band as MCUs. Similarly, the received uplink signal vector at SC (i.e., (2)) will include the MCU terms in the case of co-channel TDD.

Each element of the channel vectors, denoted as $h$, is modelled as $h = \sqrt{\beta}g$ where $\beta$ represents the large scale fading and $g$ represents the small-scale fading. The small-scale fading coefficient $g$ is modelled as Rayleigh random variable. The large-scale fading coefficient $\beta$ consist of path loss and shadowing and is given by $\beta = \frac{z}{r^{-\alpha}}$ where $r$ is the distance between the transmitter and receiver and $\alpha$ is the corresponding path loss exponent. $z$ represents log normal shadowing.

Since in general, perfect channel state information is never available at the transmitter. Therefore, to form the LS filter, the estimated channel vector can be written as:

$$\widehat{\boldsymbol{h}} = \boldsymbol{h} + \boldsymbol{h}_e \quad (3)$$

Where $\widehat{\boldsymbol{h}}$ is the estimated channel vector and $\boldsymbol{h}_e$ denotes the CEE vector, modelled as a Gaussian distributed random vector with zero mean and covariance matrix of $\sigma_e^2 \boldsymbol{I}_N$.

Given that the signals from MCU and SCU are uncorrelated, the conditional received signal covariance matrix (conditioned on channel vector $\boldsymbol{h}$, can be written as:

$$Q_i^{BS} = E\left[\boldsymbol{y}_i^{BS}(t)\left(\boldsymbol{y}_i^{BS}(t)\right)^H |\boldsymbol{h}\right]$$
$$= \sum_{b=1}^{B}\left(\sum_{k=1}^{K}P_{MCU}[\boldsymbol{h}_{ibk}^{BS-MCU}(\boldsymbol{h}_{ibk}^{BS-MCU})^H]\right) + N_0\boldsymbol{I}_N \quad (4)$$

where $(.)^H$ denotes the Hermitian of the enclosed.

Similarly, the conditional covariance matrix of the received signal at the $j^{th}$ SC in the $i^{th}$ macro-cell can be written as:

$$Q_{ij}^{SC} = E\left[\boldsymbol{y}_{ij}^{SC}(t)\left(\boldsymbol{y}_{ij}^{SC}(t)\right)^H\right]$$
$$= \sum_{b=1}^{B}\left(\sum_{s=1}^{S}P_{SCU}\left[\boldsymbol{h}_{ijbs}^{SC-SCU}(\boldsymbol{h}_{ijbs}^{SC-SCU})^H\right]\right) + N_0\boldsymbol{I}_F \quad (5)$$

where $\boldsymbol{I}_F$ is an $F \times F$ identity matrix.

For an LMMS receiver, when we have perfect channel state information, the estimated signal from the $k^{th}$ MCU at the $i^{th}$ BS and from the $s^{th}$ SCU at the $j^{th}$ SC respectively can be computed as follows (assuming $E\{x_{ij}^2\} = 1$):

$$\hat{x}_{ik}(t) = \sqrt{P_{MCU}} y_i^{BS}(t)(Q_i^{BS})^{-1}(h_{iik}^{BS-MCU})^H \quad (6)$$

$$\hat{x}_{is}(t) = \sqrt{P_{SCU}} y_{ij}^{SC}(t)(Q_{ij}^{SC})^{-1}(h_{isis}^{SC-SCU})^H \quad (7)$$

where $h_{iik}^{BS=MCU}$ is the channel vector between $k^{th}$ user in macro-cell $i$ and $i^{th}$ BS. And $h_{isis}^{SC-SCU}$ denotes the channel vector between $s^{th}$ SCU in macro-cell $i$ and $s^{th}$ SC in macro-cell $i$.

However, in practice, since incomplete channel information is available, the estimated signal from MCU can be computed in terms of received signal covariance matrix $\hat{Q}_i^{BS}$ conditioned on estimated channel vector $\hat{\boldsymbol{h}}_{iik}^{BS-MCU}$:

$$\hat{x}_{ik}(t) = \sqrt{P_{MCU}} y_i^{BS}(t)(\hat{Q}_i^{BS})^{-1}(\hat{\boldsymbol{h}}_{iik}^{BS-MCU})^H \quad (8)$$

Using (1), we have

$$\hat{x}_{ik} = \alpha P_{MCU} x_{ik} + \gamma_{ik} \quad (9)$$

Where $\alpha = \boldsymbol{h}_{iik}^{BS-MCU}(\hat{Q}_i^{BS})^{-1}(\hat{\boldsymbol{h}}_{iik}^{BS-MCU})^H$ and $\gamma_{ik}$ denote the impact of noise and interference.

Similarly, the estimated signal from SCU can be computed in terms of received signal covariance matrix $\hat{Q}_{ij}^{SC}$ conditioned on estimated channel vector $\hat{\boldsymbol{h}}_{isis}^{SC-SCU}$ as follows:

$$\hat{x}_{is}(t) = \sqrt{P_{SCU}} y_{ij}^{SC}(t)(\hat{Q}_{ij}^{SC})^{-1}(\hat{h}_{isis}^{SC-SCU})^H \quad (10)$$

Using (2), we have

$$\hat{x}_{is} = \beta P_{SCU} x_{is} + \gamma_{is} \quad (11)$$

Where $\beta = h_{isis}^{SC-SCU}(\hat{Q}_{ij}^{SC})^{-1}(\hat{h}_{isis}^{SC-SCU})^H$ and $\gamma_{is}$ denote the impact of noise and interference.

For this receiver, the signal to interference and noise (SINR) ratios when imperfect channel state information is available can be calculated for the MCU $k$ present in the $i^{th}$ macro-cell as:

$$SINR_{ik}^{UL-MCU} = \frac{P_{MCU}\alpha^2}{\hat{\boldsymbol{h}}_{iik}^{BS-MCU}(\hat{Q}_i^{BS})^{-1}Q_i^{BS}(\hat{Q}_i^{BS})^{-1}(\hat{\boldsymbol{h}}_{iik}^{BS-MCU})^H - P_{MCU}\alpha^2} \quad (12)$$

Similarly, the SINR for the SCU $s$ present in the $i^{th}$ macro-cell $i$ is given by:

$$SINR_{is}^{UL-SCU} = \frac{P_{SCU}\beta^2}{\hat{\boldsymbol{h}}_{isis}^{SC-SCU}(\hat{Q}_{ij}^{SC})^{-1}Q_{ij}^{SC}(\hat{Q}_{ij}^{SC})^{-1}(\hat{\boldsymbol{h}}_{isis}^{SC-SCU})^H - P_{SCU}\beta^2} \quad (13)$$

Using the above SINR values, we can calculate the instantaneous effective spectral efficiency of MCU $k$ and SCU $s$ present in the $i^{th}$ macro-cell as follows [1]:

$$R_{ik}^{UL-MCU} = \frac{T_{UL}}{T}\log_2(1 + SINR_{ik}^{UL-MCU}) \quad (14)$$

And

$$R_{is}^{UL-SCU} = \frac{T_{UL}}{T}\log_2(1 + SINR_{is}^{UL-SCU}) \quad (15)$$

### III. NUMERICAL RESULTS

In this section, we demonstrate the effect of channel estimation error on the performance of MIMO macro-cell versus small-cell. We consider a $3 \times 3$ grid of total **9** BSs, where each BS serves **K** MCUs. Every BS covers an area of one square kilometre over which **S** SCs are distributed on a regular grid. The MCUs are uniformly distributed over the entire cell area while one SCU is uniformly distributed within a disc of radius 40 meters around each of the SCs.

The SCUs and MCUs are associated to their closest SC and BS, even if other cell associations could provide a higher instantaneous rate. The channel model consists of small scale and large-scale fading. System parameters, path loss and shadowing parameters are specified in Table 1. We compute the Uplink sum rates of the macro and SC tier in the centre cell, averaged over 1000 different channel realizations and User distributions. By changing the fraction of the total bandwidth allocated to each tier, we obtain the UL rate regions as shown in the below figures. In Fig. 2, we assume that the macro BS consists of 20 antennas and the small cell BS consists of 1 antenna, MCUs and SCUs have one antenna each. Each macro cell has one macro BS and 20 MCUs, and has 20 small cells, each small cell has one small cell BS serving one SCU. The channel estimation error for the channel coefficients is varied from a variance of 0.01 to 0.3 for the channel between MCU and BS or SCU and SC.

As we can see from Fig. 2, that as the channel estimation error increases, spectral efficiency of both the tiers decreases drastically. However, in case of Macro users, spectral efficiency is reduced by a huge fraction, as expected, since the number of users, served by one macro

BS is large, and therefore the interference encountered is also large.

Table 1 – System parameters and path loss models.

| General system parameters | |
|---|---|
| Transmit Power | BS: 46dBm, SC: 24dBm, MCU/SCU: 23 dBm |
| Bandwidth | 20 MHz, 2 GHz centre frequency |
| Noise power spectral density | -174 dBm/Hz |
| Network Topology | B=9 macro cells, site distance=1000 m |
| UE deployment | K = 20 MCUs, 1 SCU uniformly distributed |
| Antennas | N per BS, F per SC, 1 per MCU, 1 per SCU |
| Propagation parameters | |
| Channel type | Path loss and shadowing parameters (d in meters) |
| BS-MCU/SCU | $PL_{LOS}(d) = 30.8 + 24.2 log10(d)[dB]$ <br> $PL_{NLOS}(d) = 2.7 + 42.8 log10(d)[dB]$ <br> $Pr_{LOS}(d) = \min\left(\frac{18}{d}, 1\right)\left(1 - \exp\left(-\frac{d}{72}\right)\right) + \exp\left(-\frac{d}{63}\right)$ <br> $\theta_{shadowing} = 6dB$ |
| SC-MCU/SCU | $PL_{LOS}(d) = 41.1 + 20.9 log10(d)[dB]$ <br> $PL_{NLOS}(d) = 32.9 + 37.5 log10(d)[dB]$ <br> $Pr_{LOS}(d) = 0.5 - \min\left(0.5, 5\exp\left(-\frac{156}{d}\right)\right) + \min(0.5, 5\exp\left(-\frac{d}{30}\right))$ <br> $\theta_{shadowing} = 3dB(LOS), 4dB(NLOS)$ |

For the small-cell users, change in spectral efficiency due to the channel estimation error is comparatively quite small. Therefore, each macro-BS should accurately know the channel to distinguish between the different macro users, whereas in case of small-cell BS, complexity of channel estimation techniques should be proportional to the size of the cell and the number of users being served by the small-cell BS.

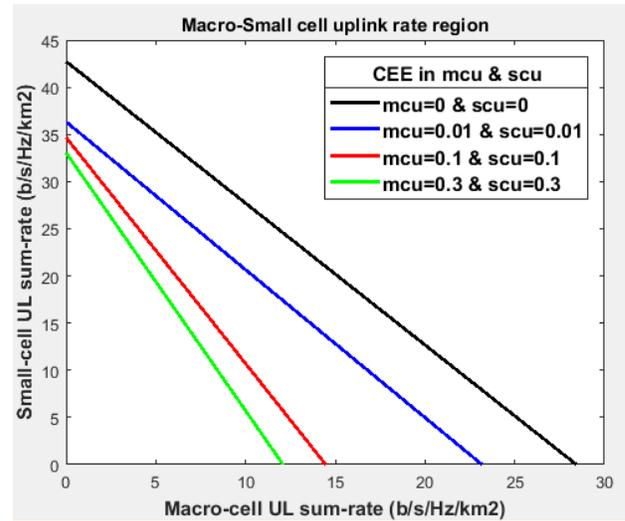

Fig. 2 – Sum Rate Analysis with 20 BS Antennas and 1 SC Antenna with channel estimation error

Fig-3 shows the sum rate analysis, considering 20 macro users and 36 small cells in one macro cell. In this plot, we have increased the number of small-cells present in one macro-cell. The idea here is to observe the effect of channel estimation error on increasing the number of small-cells present in one macro-cell. The plot shows that with increasing the number of small-cells, CEE has the same effect on the spectral efficiency of the SCUs as long as one SCU is being served by one small-cell BS.

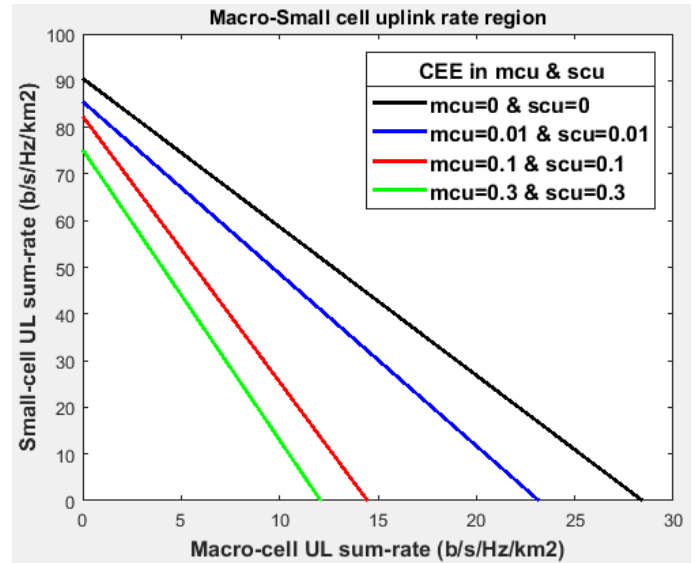

Fig. 3 - Sum Rate Analysis with 20 MCU and 36 SCU in 1 macro cell with channel estimation error

Yet another important point to be noted is that, even though CEE has comparatively less effect on small-cell users, CEE cannot be neglected as there is 25 percent drop in sum-rate spectral efficiency for a CEE of variance of 0.01.

## IV. CONCLUSION

In this paper, we have presented a heterogeneous network with massive MIMO macro tier overlaid with a dense tier of SCs. We have compared the performance of the MCUs versus SCUs based on the available channel knowledge. Simulation results showed that, in the presence of channel estimation error, the performance of macro-tier was impacted more severely than its small-cell tier counterpart. Also, even though the impact of CEE on small-cell users is somewhat small, such impact cannot be entirely ignored. It is also noteworthy that CEE has distinct impacts on various tiers of HetNets system. Hence, a tier-dependent CEE strategy must be considered for massive MIMO HetNets systems.